\documentclass[aps,prx,twocolumn,amsmath,amssymb,showpacs,superscriptaddress,notitlepage]{revtex4-1}
\usepackage{amsmath,amssymb,amsfonts,bm}
\usepackage{graphicx}
\usepackage{epstopdf}
\usepackage{dcolumn}
\usepackage{mathrsfs}
\usepackage{bbold}
\usepackage{dsfont}
\usepackage{dcolumn}
\usepackage[colorlinks=true,linkcolor=blue,citecolor=blue, urlcolor=blue,bookmarks=false]{hyperref}
\usepackage{changes}
\colorlet{Changes@Color}{red}

\begin{document}

\title{Chiral edge state coupling theory of transport in quantum anomalous Hall insulators}

\author{Rui Chen}
\affiliation{Department of Physics, Hubei University, Wuhan 430062, China}

\author{Hai-Peng Sun}
\affiliation{Institute for Theoretical Physics and Astrophysics, University of W\"urzburg, 97074 W\"urzburg, Germany}

\author{Bin Zhou}\email[]{binzhou@hubu.edu.cn}
\affiliation{Department of Physics, Hubei University, Wuhan 430062, China}
\author{Dong-Hui Xu}\email[]{donghuixu@cqu.edu.cn}
\affiliation{Department of Physics and Chongqing Key Laboratory for Strongly Coupled Physics, Chongqing University, Chongqing 400044, China}
\affiliation{Center of Quantum Materials and Devices, Chongqing University, Chongqing 400044, China}

\begin{abstract}
The quantum anomalous Hall effect is characterized by a quantized Hall resistance with a vanishing longitudinal resistance. Many experiments reported the quantization of the Hall resistance, which is always accompanied by a non-vanishing longitudinal resistance that is several k$\Omega$. Meanwhile, the non-vanishing longitudinal resistance exhibits a universal exponential decay with the increase in magnetic field. We propose that the coupling of chiral edge states, which has not been properly evaluated in the previous theories, can give rise to the non-vanishing longitudinal resistance. The coupling between the chiral edges states along the opposite boundaries can be assisted by magnetic domains or defects inside the sample bulk, which has been already identified in recent experiments. Our theory provides a potential mechanism to understand the experimental result in both magnetic topological insulator and moir\'e superlattice systems.
\end{abstract}
\maketitle
\section{Introduction}
In the last decade, the quantum anomalous Hall effect has attracted great attention due to not only  its fundamental physics interest, but also its potential applications for designing low-power consumption electronic devices~\cite{Chang2013Science,Nagaosa2010RMP,Yu2010Science,Liu2016AnnualReview,
Chang23RMPColloquium,Chi22AMProgress,Dai22PRBQuantum,
Devakul22PRXQuantum,Henk12PRLTopological,Jiang12PRBQuantum,
Jiang18PRLAntiferromagnetic,Jiang20NMConcurrence,Kawamura23NPLaughlin,
Li21PRLSpin,Li22PRLChern,Li23NSRProgress,Liu13PRLInPlane,Qin2022PRB,Okazaki22NPQuantum,Polshyn20NElectrical,
Qi16PRLHighTemperature,Qiu23NMAxion,Wang13PRLQuantum,Wang15PRLElectrically,
Wang20PRLDemonstrationa,Wang21IIntrinsic,Wang23NSRTopological,Wu14PRLTopological,
Xiao18PRLRealization,Chen23PRBSidesurfacemediated,Zhao20NTuning,Zhao22PRLZero,ChenCZSC1,ChenCZSC2}.
In many recent experiments~\cite{Chang2013Science,Checkelsky14np,Kou14prl,
Bestwick15PRL,Li2021Nature}, the Hall resistance manifests the quantization, which is accompanied by a significant longitudinal resistance ranging from 0.00013 $h/e^2$~\cite{Chang2015NatMat} to even 0.5 $h/e^2$~\cite{Checkelsky14np,Wang2018NatPhys}, even at very low temperatures. Such a scenario obviously contradicts the common belief that the quantized Hall insulator is characterized by the quantized Hall resistance and a vanishing longitudinal resistance. In the subsequent studies, the non-vanishing longitudinal resistance is interpreted as a consequence of the quasihelical edge modes~\cite{Wangjing2013PRL}, dissipative edge states~\cite{Lin2022PRB}, charge puddles in the gapped surface states~\cite{LippertzarXiv2021} or the bulk-dominated dissipation~\cite{Rosen2021arXiv,Fox18PRB,Rodenbach2021APL,Fijalkowski2021nc}, yet without consensus on its nature. Furthermore, the previous theoretical explanations for this phenomenon only apply to a rather small longitudinal resistance within several percents of $h/e^2$~\cite{He2015Physics,Wangjing2013PRL}. In the regime of quantized Hall resistance, an explanation for the occurrence of the large longitudinal resistance about 0.5 $h/e^2$~\cite{Checkelsky14np} is still absent. Moreover, the exponential decay of the longitudinal resistance with the increase in magnetic field is not well explained~\cite{Chang2013Science,Checkelsky14np,Kou14prl,
Bestwick15PRL,Li2021Nature}. A comprehensive understanding of the non-vanishing longitudinal resistance is important for  designing perfect transport of the quantum anomalous Hall effect.

\begin{figure}[h!tpb]
\centering
\includegraphics[width=0.9\columnwidth]{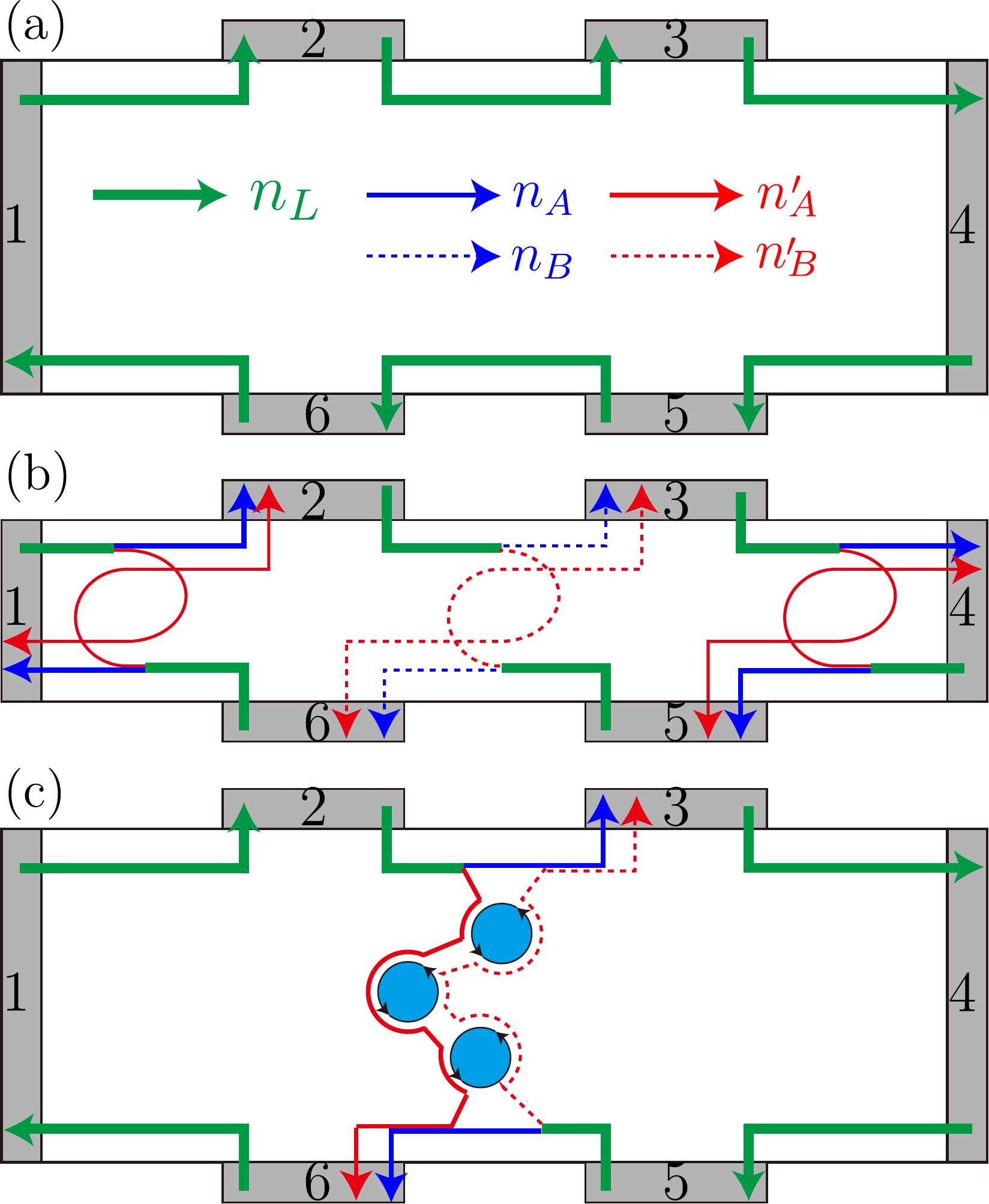}
\caption{Illustrations of the six-terminal Hall-bar setup with the propagation pattern of the chiral currents in the quantum anomalous Hall insulator. (a) The chiral current flows when there is no coupling between edge states. (b) The chiral current flows when there is a significant coupling between edge states. (c) The chiral edge current and the localized current inside the bulk when there exist magnetic domains, defects or disorders. Though the chiral edge states couple together, in the cases of (b) and (c), the transport is characterized by a quantized Hall resistance but accompanied by a non-vanishing longitudinal resistance.}
\label{fig_illustration}
\end{figure}

In this work, based on a low-energy effective model describing magnetic topological insulator films, we study the quantum transport of the quantum anomalous Hall effect~[Fig.~\ref{fig_illustration}(a)]. We demonstrate that the coupling of the chiral edge states will naturally lead to a nonzero longitudinal resistance even up to several $h/e^2$, but still remains the quantization of the Hall resistance~[Fig.~\ref{fig_illustration}(b)]. Meanwhile, the exponential decay of the longitudinal resistance with the increase in magnetic field can be well understood. At last, relevance to the recent experiments of the quantum anomalous effect is discussed.


Moreover, we reveal that the coupling of the edge states on the opposite sides can be enhanced in two ways: (i) The chiral edge states become much extended near the coercive field due to the vanishing of the magnetic gap~\cite{Zhou2008PRL,Chen2016CPB,JiangHua2014, Linder2009PRBa,Zhang2010NatPhys,LuHZ2010PRB, LiuCX2010PRB,Imura2012PRB,Takane2016JPSJ,Chen2017PRBDirac,
Wang2012PRB,Wang2013PRB,Xiao2015SciRep,Pan2015SciRep,
Schumann2018PRL,Collins2018Nature}. As a result, the longitudinal resistance should exhibit an exponential decay as a function of not only the sample size but also the strength of the external magnetic field. The former relationship is well recognized but still difficult to be observed experimentally. However, the latter one has already been observed in experiments~\cite{Checkelsky14np,Li2021Nature}, yet without explanation. (ii) The localized states inside the bulk can provide a route that bridges the edge states on the opposite sides~[Fig.~\ref{fig_illustration}(c)]. Such a mechanism induced coupling of chiral edge states has been reported in very recent experiments with the sample width up to 5 $\mu$m \cite{Qiu22PRL,Zhou23PRLConfinementInduced,Deng2022NatCom}. 

\section{Effective model and methods}
As a prototype example of the quantum anomalous Hall insulator, we consider confined quantum states around the $\Gamma$ point $\left(k_{x}= k_{y}=0\right)$ in a magnetic topological insulator thin film, which gives rise to a low-energy effective model~\cite{Lu13prl-QAH,LuHZ2010PRB,Shan2010NJP,Liu10prb,Yu2010Science}
\begin{equation}
H=H_{0}+\frac{m}{2} \tau_{0} \otimes \sigma_{z},
\end{equation}
where $m$ is the exchange field from magnetic dopants or intrinsic magnetic orderings, which can be tuned by the external magnetic field. $\tau_{0}$ is a $2 \times 2$ identity matrix. $\sigma_{z}$ is the $z$ component of the Pauli matrix. $H_{0}$ describes the topological insulator thin film without magnetization
\begin{equation}
H_{0}=\left(\begin{array}{cccc}
\frac{\Delta}{2}-B k^{2} & i \gamma k_{-} & V & 0 \\
-i \gamma k_{+} & -\frac{\Delta}{2}+B k^{2} & 0 & V \\
V & 0 & -\frac{\Delta}{2}+B k^{2} & i \gamma k_{-} \\
0 & V & -i \gamma k_{+} & \frac{\Delta}{2}-B k^{2}
\end{array}\right),
\end{equation}
with the basis being $|+\uparrow\rangle,|-\downarrow\rangle,|-\uparrow\rangle,|+\downarrow\rangle$~\cite{Yu2010Science,Sun2022arXiv}, where $|\pm \uparrow\rangle=(|t \uparrow\rangle \pm|b \uparrow\rangle) / \sqrt{2}$, $|\pm \downarrow\rangle=(|t \downarrow\rangle \pm|b \downarrow\rangle) / \sqrt{2}$, $t, b$ represent the surface states sitting on the top and bottom surfaces, and $\uparrow$, $\downarrow$ represent the spin up and down states. Here $\mathbf{k}=\left(k_{x}, k_{y}\right)$ is the two-dimensional wave vector, and $k^{2}=k_{x}^{2}+k_{y}^{2}$. The model parameters $B$ and $\Delta$ are determined by the film thickness. $\Delta$ is the hybridization of the top and bottom surface states of the thin film, which becomes irrelevant for thick films, e.g., $\mathrm{Bi}_{2} \mathrm{Se}_{3}$ thicker than $5 \mathrm{~nm}$. $\gamma=v \hbar$, with $v$ the effective velocity. $V$ corresponds to the potential distribution along the film growth direction and measures the structural inversion asymmetry between the top and bottom surfaces of
the thin film. In the following calculations, we take $\gamma=300$ meV$\cdot$nm, $V=0$ meV, $B=-300$ meV$\cdot$nm$^2$, and $\Delta=0$ meV, which is adopted from the effective parameters describing a magnetic topological insulator Cr-doped (Bi,Sb)$_2$Te$_3$~\cite{Lu13prl-QAH}. In Sec.~SII of~\cite{Supp}, we provide more calculations when non-zero $V$ and $\Delta$ are considered. The system depicts a quantum anomalous Hall phase with Chern number $C=1$ for $m>0$ and $C=-1$ for $m<0$, respectively. In the numerical calculations of the quantum transport, we discretize the effective Hamiltonian on
a simple square lattice and set the lattice constant as $a=1$ nm. The Fermi energy is taken as $E_\text{F}=3$ meV.

\begin{figure*}[htpb]
\centering
\includegraphics[width =2\columnwidth]{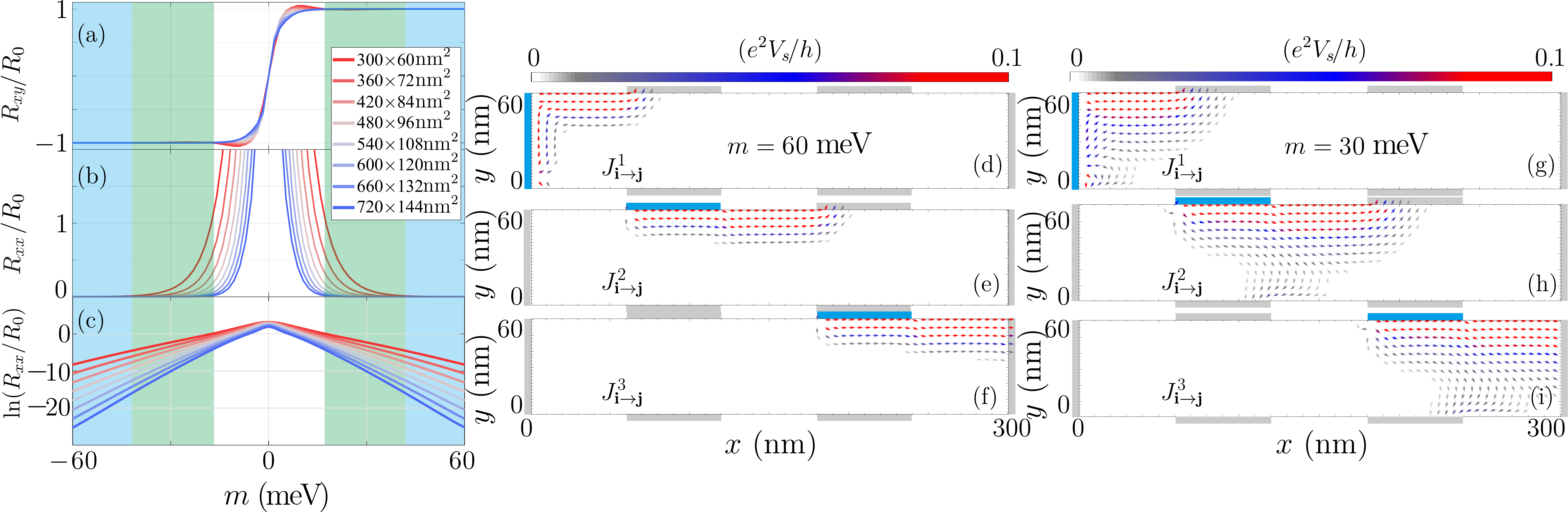}
\caption{(a) The Hall resistance $R_{xy}/R_0$, (b) longitudinal resistance $R_{xx}/R_0$, and (c) $\ln(R_{xx}/R_0)$ as functions of the exchange field $m$, respectively.  $R_0=h/e^2$ is the von Klitzing constant. The different colored curves represent different system sizes with a fixed length-width ratio as $L_x:L_y=5:1$. (d)-(f) and (g)-(i) show the nonequilibrium local current distributions injected from $s$-th terminal for $m=60$ meV and $m=30$ meV, respectively. The color of the arrows indicate the current strength marked in the color bar. }
\label{fig_clean}
\end{figure*}

We consider a standard six-terminal Hall-bar as depicted in Fig.~\ref{fig_illustration}. We investigate the transport properties of the system by using the
Landauer-B\"uttiker-Fisher-Lee formula
\cite{Landauer1970Philosophical,Buttiker1988PRB,
Fisher1981PRB} and the recursive Green's
function method \cite{Mackinnon1985Zeitschrift,Metalidis2005PRB}. The current flowing into terminal $p$ is given by $
I_{p}=\frac{e^{2}}{h}\sum_{q\neq p}T_{pq}\left( E_\text{F}\right) \left(
V_{p}-V_{q}\right)$~\cite{Landauer1970Philosophical,Buttiker1988PRB}, where $T_{pq}$ depicts the transmission coefficient from electrode $p$ to $q$, and $V_i$ corresponds to the voltage of lead $i$ shown in Fig.~\ref{fig_clean}(a). The longitudinal and Hall resistances are given by $R_{xx}=\left(V_3-V_2\right)/I$ and $R_{xy}=\left(V_6-V_2\right)/I$, respectively (see Sec.~SI of~\cite{Supp} for more details). The linear conductance can be obtained by the transmission coefficient $T_{pq}$ from terminal $p$ to terminal $q$, where $T_{pq}=$ Tr$\left[ \Gamma^p G^r \Gamma^q G^a \right]$ ($p, q=1,2,\ldots,6$ and $p\neq q$). The linewidth function is defined as $\Gamma^{p}(\mu)=i\left[ \Sigma_{p}^{r}-\Sigma_{p}^{a}\right]$ with $\Sigma_{p}^{r/a}$ being the retarded/advanced self-energy at the terminal $p$, and the Green's functions $G^{r/a}$ are calculated from $G^{r}=\left( G^a\right)^\dag=\left[E_\text{F}-H^\text{C}-\sum_{p}\Sigma_{p}^{r} \right]^{-1}$, where $E_\text{F}$ is the Fermi energy and $H^\text{C}$ is the Hamiltonian matrix of the central scattering region. To calculate the local current distribution, a small external bias $V_s$ is applied between the $s$-th terminal and all the other terminals. The nonequilibrium local
current distribution between sites $\mathbf{{i}}$ and $\mathbf{{j}}$ can be obtained from the following formula~\cite{Jiang2009PRB}
\begin{equation}
J_{\mathbf{{i}\rightarrow{j}}}^{s}=\frac{2e^{2}}{h}\text{Im} \left[  \sum
_{\alpha,\beta}{H^\text{C}_{\mathbf{{i} \alpha,{j}\beta}}G^{n,s}_{\mathbf{{j}\beta
			,{i}\alpha}}}\right]    V_s   \text{,}%
\end{equation}
where
$G^{n,s}=G^{r}\Gamma^s G^{a}$ is the electron correlation function.


\section{Size and magnetism dependent chiral edge state coupling}
To gain insight into the coupling of chiral edge states, we consider a varying exchange field $m$ and calculate the Hall and longitudinal resistance for various system sizes, which are shown in Figs.~\ref{fig_clean}(a) and \ref{fig_clean}(b). When $\left|m\right|$ is large enough (the light blue region), we observe a perfect quantum anomalous Hall effect with the quantized Hall resistance and vanishing longitudinal resistance, which is not sensitive to the sizes we consider. As $\left|m\right|$ decreases (the light cyan region), an interesting phenomenon occurs. Though $R_{xy}$ remains the quantized value of $h/e^2$, $R_{xx}$ dramatically increases with the decrease in the sample size, as shown in the light cyan shaded regime. By further reducing $\left|m\right|$~(the white region), $R_{xy}$ deviates from the quantized plateau.

For about $m=60$ meV (the light blue region) with the system size being $L_x\times L_y=300\times60$ nm$^2$, the corresponding transmission matrix is given by
\begin{equation}
T=%
\begin{pmatrix}
0.000&1.000&0.000&0.000&0.000&0.000\\
0.000&0.000&1.000&0.000&0.000&0.000\\
0.000&0.000&0.000&1.000&0.000&0.000\\
0.000&0.000&0.000&0.000&1.000&0.000\\
0.000&0.000&0.000&0.000&0.000&1.000\\
1.000&0.000&0.000&0.000&0.000&0.000\\
\end{pmatrix}%
.
\end{equation}
The above transmission matrix depicts the well-defined quantum anomalous Hall effect with a dissipationless chiral current propagating along the system boundary. From the above transmission matrix, we have $R_{xy}=1.000~h/e^2$ and $R_{xx}=0.000~h/e^2$. The scenario is illustrated more clearly by calculating the current distribution shown in Figs.~\ref{fig_clean}(d)-\ref{fig_clean}(f), where the localization length is smaller compared to the system size.

For about $m=30$ meV (the light cyan region) with the same sample size, the transmission matrix has the form
\begin{equation}
T=%
\begin{pmatrix}
0.000&0.888&0.001&0.000&0.000&0.002\\
0.002&0.000&0.910&0.001&0.002&0.088\\
0.000&0.000&0.000&0.888&0.113&0.002\\
0.000&0.000&0.002&0.000&0.888&0.001\\
0.001&0.002&0.088&0.002&0.000&0.910\\
0.888&0.113&0.002&0.000&0.000&0.000\\
\end{pmatrix}%
.
\end{equation}
Though each element of the above transmission matrix deviates from integer values, but it still gives rise to a nearly quantized Hall resistance $R_{xy}=0.994~h/e^2$ and a significant longitudinal resistance $R_{xx}=0.101~h/e^2$. It is noted that the summation of each column of the transmission matrix may slightly exceed one by a small amount (on the order of 0.001 for $m=30$ meV), which is attributed to the band broadening effect in the numerical calculations.

\begin{figure*}[ht]
\centering
\includegraphics[width =1.7\columnwidth]{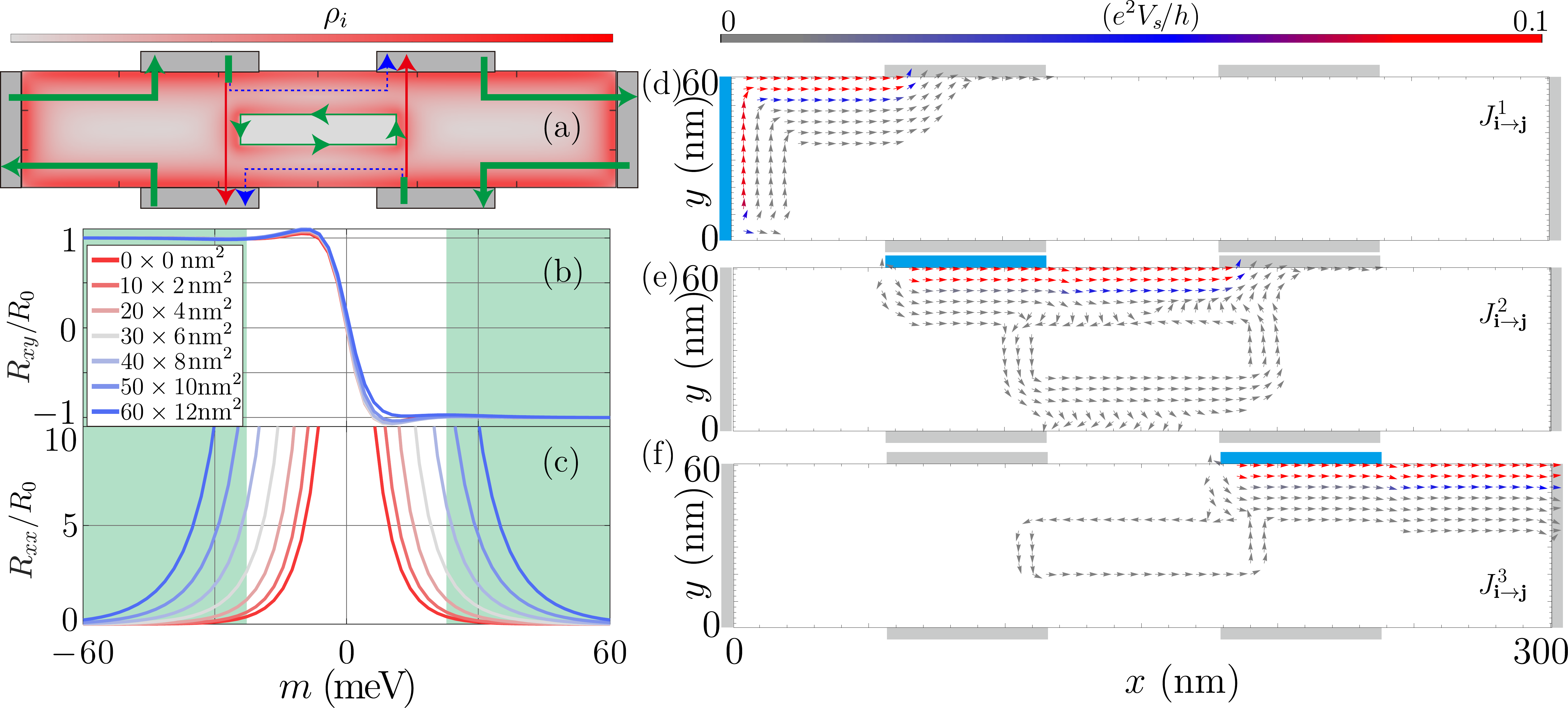}
\caption{(a) Illustration of the six-terminal Hall bar with a rectangle domain. The color depicts the local density of states defined as $\rho_{\mathbf{i}}=-\frac{1}{\pi}\text{Im}G^r_{\mathbf{ii}}$. (b) The Hall resistance $R_{xy}/R_0$ and (c) longitudinal resistance $R_{xx}/R_0$ as a function of $m$ for different domain size $L_x^d : L_y^d=5:1$, respectively. The system size is $L_x\times L_y=300\times 60$ nm$^2$. (d)-(f) The nonequilibrium local current distributions injected from $s$-th terminal (colored by blue) for $L_x^d \times L_y^d=60\times 12$ nm$^2$ and $L_x \times L_y=300\times 60$ nm$^2$ with $m=60$ meV.}
\label{fig_domain}
\end{figure*}

To explain this, let us first neglect the small matrix elements and rewrite the above transmission matrices in the following form
\begin{equation}
T\approx%
\begin{pmatrix}
0&n_A&0&0&0&0\\
0&0&n_B&0&0&n_B'\\
0&0&0&n_A&n_A'&0\\
0&0&0&0&n_A&0\\
0&0&n_B'&0&0&n_B\\
n_A&n_A'&0&0&0&0\\
\end{pmatrix}%
.
\end{equation}
Note we have $n_A+n_A'=1.001$ and $n_B+n_B'=0.998$ for $m=30$ meV. We observe that $n_A+n_A'\approx n_B+n_B' \approx1$ is satisfied as long as $m$ falls in the light cyan region shown in Fig.~\ref{fig_clean}. Note there is a slight difference between $n_A^{(\prime)}$ and $n_B^{(\prime)}$, which is attributed to the distinct propagating patterns when electrons are injected from different terminals [see Fig.~\ref{fig_illustration}(b) and Figs.~\ref{fig_clean}(g)-\ref{fig_clean}(i)]. Moreover, the difference is negligible in large systems due to the vanishing of the coupling between the chiral edge states on opposite boundaries. Accordingly, the Hall and longitudinal resistances are given by
\begin{align}
R_{xy}=\frac{1}{n_A+n_A'},
R_{xx}=\frac{n_B'}{ n_B(n_A+n_A')}.
\end{align}
Near $m=30$ meV, the localization length of the edge states is comparable to the system size, and the edge states on the opposite sides couple together. Therefore $n_A$, $n_A'$, $n_B$, and $n_B'$ deviate from the integer values $1$ and $0$, respectively. The well-quantized Hall resistance is mainly contributed from the edge channel, but accompanied by a non-zero $R_{xx}$. Further decreasing $m$, the localization length of the chiral edge states grows, resulting in a large longitudinal resistance.

Above, we reveal that the emergency of the non-vanishing longitudinal resistance is attributed to the exchange field dependent localization length of chiral edge states. When the localization length is comparable to the sample size, the coupling of the edge states is captured by a finite-size gap $E_g$, which is proportional to $e^{-mL_y/2\gamma}$, where $L_y$ is the sample width (see Sec.~SII of~\cite{Supp} for more details). This conclusion is confirmed in Fig.~\ref{fig_clean}(c), which shows a linear relationship $\ln (R_{xx}/R_0)\sim m$. On the other hand, the formula indicates that the coupling of edge states is significantly enhanced near the coercive filed where the exchange field gap $m$ is vanishing. We would like to point out that the exponential behavior of the longitudinal resistance with the external magnetic field $\ln (R_{xx}/R_0)\sim B_z$ had been observed in various experiments~\cite{Li2021Nature,Checkelsky14np,Chang2013Science,
Mogi17sa,Deng2022NatCom,Grauer17PRL,Chang16PRL,YangF2015PRL}, where $B_z$ is the strength of the external perpendicular magnetic field. These observations agree well with our result as $m$ is proportional to the strength of $B_z$.

\section{Domain enhanced chiral edge state coupling}
In this part, we combine the effect of domain walls into our theory since domains are pretty common in magnetic topological insulators. To simulate this, we consider a hollow geometry as shown in Fig.~\ref{fig_domain}(a), which generates clockwise chiral currents near the outside boundary, and anti-clockwise chiral currents inside the bulk. Figures~\ref{fig_domain}(b)-\ref{fig_domain}(c) show Hall resistance and longitudinal resistances as a function of $m$ for different domain size. In the light cyan regime, the Hall resistance remains quantized, while the longitudinal resistance grows rapidly with increasing the domain size. To understand this, we plot the nonequilibrium local current distribution $J_{\mathbf{{i}\rightarrow{j}}}^{s}$ in Figs.~\ref{fig_domain}(d)-\ref{fig_domain}(f) for a domain of $60\times 12$ nm$^2$. Here, the current is injected from the $s$-th terminal. Clearly, the existence of the domain could induce the substantial coupling of edge states. Actually, in a recent experiment on a magnetic topological insulator~\cite{Qiu22PRL}, the edge state coupling is mediated by the domain walls, which results in a large resistance fluctuations near the coercive field. We also would like to point out that defects or impurities in the bulk can cause the coupling of edge states (see Sec.~SII of~\cite{Supp} for more details).



\begin{figure}[t!]
\centering
\includegraphics[width =\columnwidth]{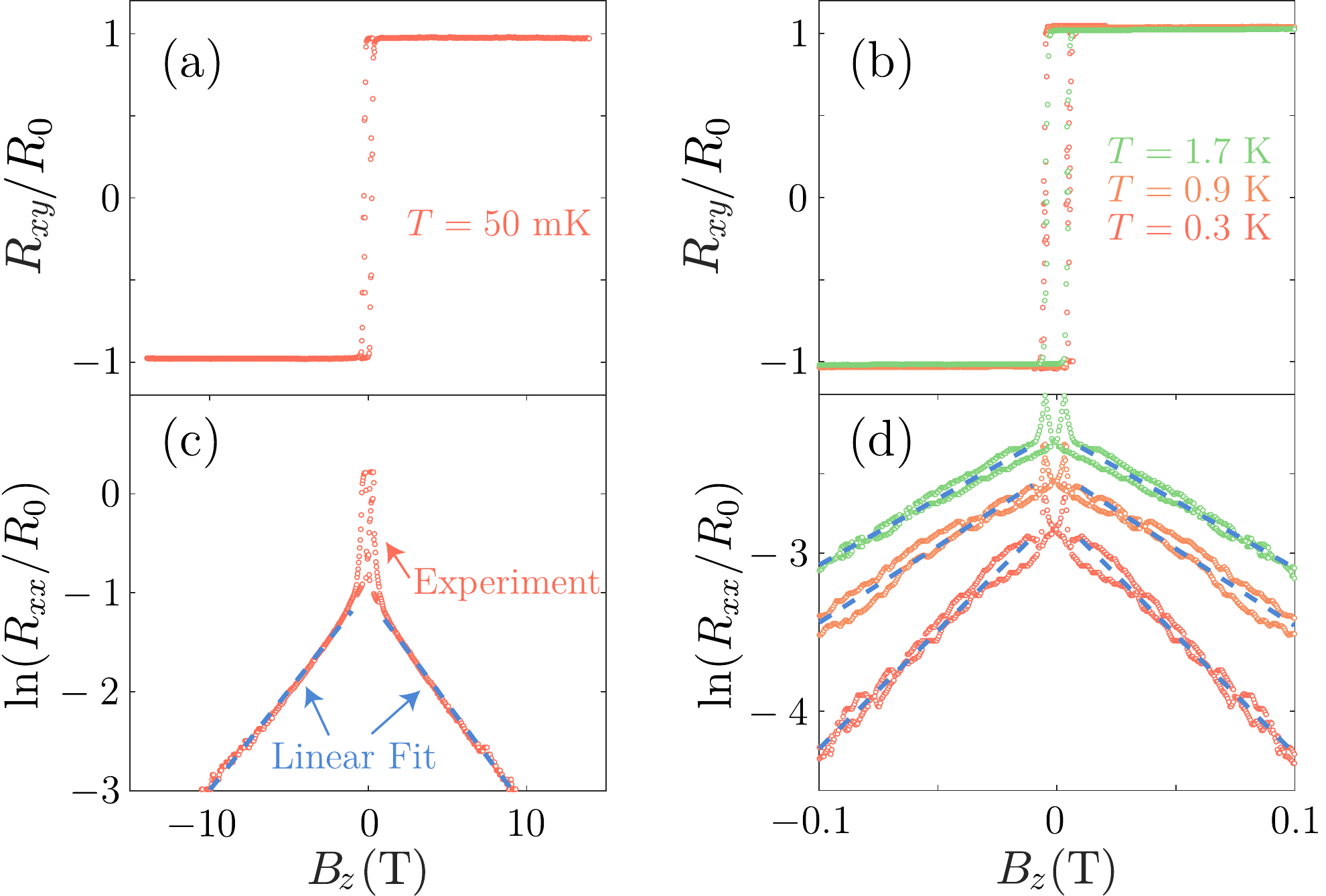}
\caption{Experimental data of $R_{xy}/R_0$ and $\ln(R_{xx}/R_0)$. (a, c) are adopted from Ref.~\cite{Checkelsky14np} and (b, d) are adopted from Ref.~\cite{Li2021Nature}. The blue dashed lines in (c) and (d) are the fitting curves with the form $\ln(R_{xx}/R_0)=a+b B_z$, with $a$ and $b$ being the fitting parameters.}
\label{fig_experiment}
\end{figure}

\section{Conclusions and discussions.}
In summary, we reveal how the coupling of edge states determines the transport properties of quantum anomalous Hall insulators. We find that the coupling of the chiral edge states is responsible for the emergence of the quantized Hall resistance. Moreover, in the quantized Hall resistance regime, we also show that the longitudinal resistance decays exponentially with the increase in magnetic field can be well explained by the chiral edge state coupling theory. In addition, the mechanism in this work is distinct from a prior work~\cite{ChenCZ2017PRB}, which also explored the non-zero longitudinal resistance induced by domain walls in the regime of quantized Hall resistance.

Our findings are universal and can be applied to various quantum anomalous Hall insulators in experiments. In Fig.~\ref{fig_experiment}, we present a comparison between experimental data and the fitting curves obtained by the formula we found. Most saliently, although the experimental data is from two different systems: one is the magnetically doped topological insulator~\cite{Checkelsky14np} and the other is transition metal dichalcogenide moir\'e heterobilayers~\cite{Li2021Nature}, the fitting curves agree well with it.


Moreover, in Secs.~SII E-F of~\cite{Supp}, we show that the above conclusions are robust when non-zero $V$ and $\Delta$ are considered. We confirm that the longitudinal resistance is more significant in thick films with a small $\Delta$, which had been reported in the previous experiments~\cite{Kou14prl}. We also consider different shapes of domain with different locations and all the cases show that the longitudinal resistance is further enhanced~(see Sec.~SII B of~\cite{Supp}). We adopt a percolation-type random lattice to show that sample defects can also enhance the longitudinal resistance in the quantized Hall regime~(see Sec.~SII C of~\cite{Supp}). Furthermore, we show that the quantized Hall resistance is robust against weak disorder, though accompanied by a non-vanishing longitudinal resistance~(see Sec.~SII G of~\cite{Supp}).

\section*{Acknowledgments}
D.-H.X. was supported by the NSFC (under Grant Nos. 12074108 and 12147102) and the Natural Science Foundation of Chongqing (Grant No. CSTB2022NSCQ-MSX0568). B.Z. was supported by the NSFC (under Grant No. 12074107), the program of outstanding young and middle-aged scientific and technological innovation team of colleges and universities in Hubei Province (under Grant No. T2020001) and the innovation group project of the natural science foundation of Hubei Province of China (under Grant No. 2022CFA012). H.-P.S. was supported by the DFG (SPP1666 and SFB1170 ``ToCoTronics''), the W\"urzburg-Dresden Cluster of Excellence ct.qmat, EXC2147, Project-id 390858490, and the Elitenetzwerk Bayern Graduate School on ``Topological Insulators''.

\bibliographystyle{scpma}
\bibliography{refs-transport}


\end{document}